\documentclass[conference]{IEEEtran}
\IEEEoverridecommandlockouts
\usepackage{cite}
\usepackage{amsmath,amssymb,amsfonts}
\usepackage{algorithmic}
\usepackage{graphicx}
\usepackage{subcaption}  
\usepackage{textcomp}
\usepackage{xcolor}
\usepackage{booktabs}
\usepackage{multirow}
\usepackage{placeins}
\def\BibTeX{{\rm B\kern-.05em{\sc i\kern-.025em b}\kern-.08em
    T\kern-.1667em\lower.7ex\hbox{E}\kern-.125emX}}

\begin{document}

\title{SOPF-Based Adaptive Droop Control for Hybrid AC--HVDC Grids Under Offshore Wind Uncertainty
}

\author{\IEEEauthorblockN{Hongjin Du}
\IEEEauthorblockA{\textit{Department of Electrical Sustainable Energy} \\
\textit{Delft University of Technology}\\
Delft, Netherlands \\
H.Du@tudelft.nl}
\and
\IEEEauthorblockN{Aleksandra Lekić}
\IEEEauthorblockA{\textit{Department of Electrical Sustainable Energy} \\
\textit{Delft University of Technology}\\
Delft, Netherlands \\
A.Lekic@tudelft.nl}
}

\maketitle
\begin{abstract}
The integration of massive offshore wind into hybrid AC-HVDC grids demands robust DC voltage regulation, yet conventional fixed-gain droop controllers struggle under severe stochastic volatility. This paper bridges the gap between system-level economic dispatch and converter-level control by proposing a novel Stochastic Optimal Power Flow (SOPF)-based adaptive droop framework. Rather than relying on heuristic or reactive tuning, wind forecast uncertainty is modeled using a zone-wise Beta distribution that accurately captures the heteroscedastic nature of wind errors across low, mid, and high power regimes. By leveraging Polynomial Chaos Expansion (PCE) within a chance-constrained SOPF, the system's stochastic states are formulated analytically. Crucially, the optimal adaptive droop gain is extracted directly from the first-order PCE coefficients via a Jacobian-free sensitivity analysis, embedding statistical voltage-security guarantees directly into the local converter control. Validation on a 4-terminal AC-HVDC system demonstrates that scenario-adaptive gains significantly outperform standard fixed-coefficient approaches, effectively minimizing active-power tracking errors during extreme wind disturbances.
\end{abstract}


\section{Introduction}
\label{sec:introduction}

\IEEEPARstart{T}{he} rapid expansion of offshore wind capacity has positioned high-voltage direct current (HVDC) transmission as a key enabler for long-distance bulk power delivery and asynchronous grid interconnection~\cite{VanHertem2010HVDC,Rodriguez2017MTDC}. In Europe, the planned North Sea offshore grid and its onshore HVDC backbone are expected to support a target of at least 450~GW of offshore wind by 2050~\cite{Yurtseven2025SOPF}, reinforcing multi-terminal HVDC (MTDC) systems built on voltage source converters (VSCs) as the predominant architecture for hybrid AC--HVDC grids~\cite{Lekic2020Init}.

Coordinated DC voltage regulation and active power sharing constitute the central control challenge in MTDC systems, with decentralized strategies broadly classified into three categories~\cite{Raza2025Review,Vrana2013Classification}. Master--slave control assigns a single station to maintain the DC voltage while the others operate in constant-power mode, which is simple but lacks redundancy since loss of the master triggers system-wide voltage collapse. Voltage margin control mitigates this by defining backup regulators that take over at power limits, at the cost of additional tuning complexity~\cite{Santos2022VoltageMargin}. DC voltage droop control distributes regulation across multiple stations without inter-station communication, offering redundancy and scalability~\cite{Beerten2013Analysis}, and has become the most widely adopted mechanism for MTDC grids integrating offshore wind~\cite{Raza2022Variable}. In its conventional form, droop control imposes an affine relationship between converter power and local DC voltage through a fixed coefficient calibrated at a nominal operating point; however, the stochastic variability introduced by offshore wind drives the system away from this nominal condition on a sub-hourly basis, so that fixed coefficients often produce excessive voltage deviations, suboptimal dispatch, and uneven converter stress following large fluctuations or contingencies~\cite{Wang2020DeviationDroop}.

Adaptive and nonlinear droop schemes have therefore been extensively investigated, including deviation-dependent formulations that recalibrate gains from real-time voltage or power excursions~\cite{Wang2020DeviationDroop,Prabhakaran2018Nonlinear}, headroom-based methods exploiting remaining converter capacity~\cite{Raza2022Variable,Li2017Enhanced}, reference-power schemes that retune set-points under large operating shifts~\cite{Wang2023Adaptive}, and look-up-table approaches precomputed for $N$-1 contingency sets~\cite{Alavi2022LookupTable}. These methods enhance robustness but adjust coefficients reactively to measured deviations rather than deriving them from an explicit probabilistic uncertainty model, so the resulting parameters carry no direct statistical voltage-security guarantee, and the tuning is largely heuristic and decoupled from system-level dispatch. This gap suggests that an optimization-based dispatch framework, properly coupled with converter-level control, could supply a principled basis for selecting droop coefficients that jointly reflect economic efficiency and probabilistic voltage security.

Optimal power flow (OPF) under uncertainty offers such a framework and, depending on the information available on uncertain inputs, is cast as robust OPF~\cite{Lee2024RobustACOPF}, distributionally robust OPF~\cite{Xie2018DROPF}, or stochastic OPF (SOPF)~\cite{Roald2023Review}. Within the SOPF paradigm, polynomial chaos expansion (PCE) has emerged as a non-intrusive technique for propagating continuous random variables through the nonlinear power-flow equations with controllable accuracy~\cite{Muhlpfordt2019CCOPF, David2020Efficient}, and the current--voltage (IV) formulation enables chance constraints to be expressed as second-order cone constraints~\cite{VanAcker2022IVRgPC}. The framework has recently been extended to hybrid AC/DC grids with non-Gaussian uncertainty and renewable curtailment, with demonstrated scalability to large systems~\cite{Yurtseven2025SOPF}. These formulations, however, typically adopt fixed distributional assumptions that are uncalibrated to actual forecast-error statistics across distinct wind regimes, and the probabilistic information remains confined to the dispatch layer without being propagated to the control level.

This paper addresses that gap by proposing an uncertainty-aware adaptive droop framework for a hybrid AC--HVDC system. Wind forecast uncertainty is represented by a Beta distribution whose shape parameters are fitted from historical forecast error statistics partitioned into three power-dependent operating zones, thereby reflecting the heteroscedastic nature of offshore wind errors rather than assuming a single fixed distribution. This characterization is embedded in a PCE-based chance-constrained SOPF in the IV space; the first-order expansion coefficients of converter power and DC voltage are then combined in closed form to yield an adaptive droop gain, bypassing the explicit formation and inversion of the Jacobian. The main contributions are as follows:
\begin{enumerate}
  \item A unified framework coupling system-level stochastic dispatch with converter-level droop control, such that the droop coefficient inherits the probabilistic voltage-security guarantee encoded in the SOPF chance constraints.
  \item A zone-wise Beta parameterization fitted from historical forecast error statistics, providing a heteroscedastic, data-driven uncertainty input that outperforms fixed-distribution assumptions commonly used in PCE-SOPF studies.
  \item A sensitivity-based droop extraction rule derived directly from the first-order PCE coefficients, bypassing explicit Jacobian inversion and yielding an interpretable adaptive gain that varies continuously with the stochastic operating point.
\end{enumerate}

The rest of this paper unfolds as follows: Section~\ref{sec:methodology} establishes the theoretical framework, detailing the zone-wise wind uncertainty model, the PCE-driven chance-constrained SOPF, and the derivation of the sensitivity-based droop gain. Section~\ref{sec:results} then validates the proposed control strategy through numerical case studies on a hybrid AC--HVDC test system. Finally, Section~\ref{sec:conclusion} summarizes the key findings and outlines future research directions.

\section{Methodology}
\label{sec:methodology}

\subsection{Zone-Wise Beta Parameterization of Wind Uncertainty}
\label{subsec:wind}

Offshore wind forecast errors exhibit pronounced heteroscedasticity: the statistical characteristics of the errors depend strongly on the wind farm's operating regime, with cut-in, ramp, and near-rated conditions each displaying distinct mean biases and variances. Representing this variability by a single distribution across all operating points, as commonly assumed in existing SOPF studies, therefore either under- or over-estimates the true uncertainty depending on the instantaneous forecast level. To capture this behavior in a tractable manner while remaining compatible with the Jacobi orthogonal polynomial basis required by the PCE framework, we adopt a zone-wise Beta parameterization. The dataset employed in this study is sourced from the IEEE DataPort “Hybrid Energy Forecasting and Trading Competition,” maintaining strict consistency with the empirical data utilized in our prior work \cite{heft2023, du2025opf}. 

Let $p_{\text{pred}}$ denote the point forecast of the wind farm output and $p_{\max}$ its rated capacity. The normalized forecast $\tilde{p} = p_{\text{pred}}/p_{\max} \in [0,1]$ is partitioned into three operating zones:
\begin{equation}
\label{eq:zones}
z(\tilde{p}) =
\begin{cases}
\text{Low}, & \tilde{p} < \tau_1, \\
\text{Mid}, & \tau_1 \leq \tilde{p} < \tau_2, \\
\text{High}, & \tilde{p} \geq \tau_2,
\end{cases}
\end{equation}
with thresholds $\tau_1 = 0.3$ and $\tau_2 = 0.7$ selected to approximately separate the cut-in (low), ramp (mid), and near-rated (high) regions of a typical offshore turbine power curve. For each zone $z$, the mean $\mu_z$ and standard deviation $\sigma_z$ of the normalized forecast error are estimated from a large archive of historical forecast--realization pairs. Given a forecast $p_{\text{pred}}$ falling in zone $z$, the effective mean of the stochastic power output is shifted by the zone-specific bias:
\begin{equation}
\label{eq:mu_eff}
\mu_{\text{eff}} = \tilde{p} + \mu_z,
\end{equation}
and the Beta shape parameters are obtained by moment matching:
\begin{subequations}
\label{eq:beta_ab}
\begin{align}
    \alpha &= \mu_{\text{eff}} K, \label{eq:beta_alpha} \\
    \beta &= (1-\mu_{\text{eff}}) K, \label{eq:beta_beta} \\
    K &= \frac{\mu_{\text{eff}}(1-\mu_{\text{eff}})}{\sigma_z^2} - 1. \label{eq:beta_K}
\end{align}
\end{subequations}
The resulting $p_w \sim \text{Beta}(\alpha,\beta)$ provides a bounded, non-Gaussian, forecast-conditioned representation of the wind power uncertainty that adapts automatically to the prevailing operating zone. The zone-specific statistics used in this work are summarized in Table~\ref{tab:zone_stats}.
\begin{table}[t]
\centering
\caption{Zone-wise forecast error statistics used in the Beta parameterization.}
\label{tab:zone_stats}
\renewcommand{\arraystretch}{1.15}
\begin{tabular}{cccc}
\toprule
Zone & Range of $\tilde{p}$ & $\mu_z$ & $\sigma_z$ \\
\midrule
Low  & $[0,\,0.3)$   & 0.011 & 0.035 \\
Mid  & $[0.3,\,0.7)$ & 0.031 & 0.081 \\
High & $[0.7,\,1.0]$ & 0.013 & 0.100 \\
\bottomrule
\end{tabular}
\end{table}
Two properties are worth highlighting: the Beta support on $[0,1]$ is physically consistent with the bounded nature of normalized wind power, in contrast to Gaussian surrogates that admit unrealistic negative or super-rated realizations; and the state-dependent variance reflects the fact that forecast errors are typically small in the cut-in and near-rated regions where the power curve is locally flat, and substantially larger in the ramp region where small wind-speed errors translate into large power deviations.

\subsection{PCE-Based Chance-Constrained SOPF}
\label{subsec:sopf}

The stochastic dispatch problem is formulated as a chance-constrained OPF in the IV space of the hybrid AC--HVDC grid, with uncertainty propagated through a finite polynomial chaos expansion. Let $\boldsymbol{\xi} = (\xi_1,\ldots,\xi_{N_\omega})$ denote the vector of independent standardized random germs associated with the $N_\omega$ wind farms, each governed by the Beta distribution of~\eqref{eq:beta_ab}. Any square-integrable random state or decision variable $x(\boldsymbol{\xi})$ of the system admits the truncated PCE representation
\begin{equation}
\label{eq:pce_expansion}
x(\boldsymbol{\xi}) \approx \sum_{k=0}^{K} \hat{x}_k\,\Phi_k(\boldsymbol{\xi}),
\end{equation}
where $\{\Phi_k\}$ is the multivariate Jacobi polynomial basis orthogonal with respect to the joint Beta density, $\hat{x}_k$ are deterministic coefficients, and the total number of basis functions is $K+1 = (N_\omega + d)!/(N_\omega!\,d!)$ for maximum degree $d$. An expansion degree $d=2$ is adopted throughout, which has been shown to give accurate results for nonlinear AC power-flow equations~\cite{Muhlpfordt2019CCOPF,Yurtseven2025SOPF}.

The chance-constrained SOPF minimizes the expected generation cost subject to stochastic power-balance and security constraints:
\begin{subequations}
\label{eq:sopf}
\begin{align}
\min_{\mathbf{x}}\ & \mathbb{E}\!\left[\sum_{i \in \mathcal{G}} C_i\big(P_{g,i}(\boldsymbol{\xi})\big)\right], \label{eq:sopf_obj}\\
\text{s.t.}\ & \mathbf{g}(\mathbf{x},\boldsymbol{\xi}) = 0, \label{eq:sopf_balance}\\
& \mathbb{P}\!\left(V^{\min}_{\text{dc},n} \leq V_{\text{dc},n}(\boldsymbol{\xi}) \leq V^{\max}_{\text{dc},n}\right) \geq 1-\epsilon, \label{eq:sopf_vdc}\\
& \mathbb{P}\!\left(h_\ell(\mathbf{x},\boldsymbol{\xi}) \leq 0\right) \geq 1-\epsilon, \label{eq:sopf_rest}
\end{align}
\end{subequations}
where $\mathbf{g}(\cdot)=0$ collects the AC, DC, and converter power-flow equations in the IV formulation, \eqref{eq:sopf_vdc} enforces DC bus voltage limits at every DC node $n$ with confidence level $1-\epsilon$, and \eqref{eq:sopf_rest} groups the remaining chance-constrained operational limits such as converter ratings and branch currents.

Substituting the PCE representation~\eqref{eq:pce_expansion} into~\eqref{eq:sopf_balance} and applying Galerkin projection onto each basis function $\Phi_k$ yields an equivalent deterministic system of equations in the coefficient space, with the number of equations scaling linearly in $K+1$. The chance constraints~\eqref{eq:sopf_vdc}--\eqref{eq:sopf_rest} are reformulated using the analytical mean and variance expressed directly from the PCE coefficients:
\begin{subequations}
\label{eq:pce_moments}
\begin{align}
    \mathbb{E}[x] &= \hat{x}_0, \label{eq:pce_mean} \\
    \mathrm{Var}[x] &= \sum_{k=1}^{K} \gamma_k \hat{x}_k^2, \label{eq:pce_var}
\end{align}
\end{subequations}
where $\gamma_k = \mathbb{E}[\Phi_k^2]$ are the known basis-norm constants. Combined with a concentration inequality, this produces tractable second-order cone constraints in the IV formulation~\cite{VanAcker2022IVRgPC}. The overall problem is solved once, in the space of PCE coefficients, using an interior-point solver; beyond the expected optimal dispatch encoded in $\hat{x}_0$, the solution delivers the full set of higher-order coefficients $\hat{x}_k$ ($k \geq 1$) which underpins the droop extraction developed next.

\subsection{Sensitivity-Based Adaptive Droop Extraction}
\label{subsec:droop}
The central observation underlying the proposed droop extraction is that the first-order PCE coefficients admit a direct sensitivity interpretation. For a state variable $x(\boldsymbol{\xi})$ expanded as in~\eqref{eq:pce_expansion}, differentiating with respect to the $i$-th random germ at the nominal point yields:
\begin{equation}
\label{eq:sensitivity}
\left.\frac{\partial x}{\partial \xi_i}\right|_{\boldsymbol{\xi}=\boldsymbol{\xi}_0} = \hat{x}_{1,i}\,\Phi_{1,i}'(\boldsymbol{\xi}_0),
\end{equation}
where $\hat{x}_{1,i}$ denotes the first-order PCE coefficient associated with $\xi_i$ and $\Phi_{1,i}$ is the corresponding first-degree Jacobi polynomial with known derivative. The first-order coefficient is therefore proportional to the partial derivative of $x$ with respect to the underlying uncertainty source. In contrast to classical sensitivity analysis, which requires explicit formation of the hybrid AC--DC power flow Jacobian matrix $\mathbf{J} = \partial\mathbf{g}/\partial\mathbf{x}$ and its numerical inversion at each operating point, the Galerkin-projected SOPF inherently embeds this information into the PCE coefficients, so that a single solve of~\eqref{eq:sopf} provides all state-variable sensitivities as a by-product.

Consider a droop-controlled converter station with active power injection $P_{\text{vsc}}$ and DC terminal voltage $V_{\text{dc}}$, both of which are random variables under wind uncertainty. The local droop coefficient is defined as the magnitude of the sensitivity of $P_{\text{vsc}}$ to $V_{\text{dc}}$, which in the PCE coefficient space takes the form
\begin{equation}
\label{eq:k_single}
\tilde{k} = \left|\frac{\partial P_{\text{vsc}}}{\partial V_{\text{dc}}}\right|
\approx \left|\frac{\hat{p}_{1,i}}{\hat{v}_{1,i}}\right|,
\end{equation}
when the system is driven by a single uncertainty source $\xi_i$. For $N_\omega$ independent wind sources, the converter is subject to multiple first-order sensitivities $(\hat{p}_{1,i},\hat{v}_{1,i})$, $i \in \{1,\ldots, N_\omega\}$. A single scalar droop gain is then obtained by a voltage-sensitivity-weighted aggregation:
\begin{subequations}
\label{eq:k_multi_group}
\begin{align}
    \tilde{k} &= \sum_{i=1}^{N_\omega} w_i \left|\frac{\hat{p}_{1,i}}{\hat{v}_{1,i}}\right|, \label{eq:k_opt} \\
    w_i &= \frac{|\hat{v}_{1,i}|}{\sum_{j=1}^{N_\omega}|\hat{v}_{1,j}|}, \label{eq:k_weight}
\end{align}
\end{subequations}
which simplifies to the compact form:
\begin{equation}
\label{eq:k_closed}
\tilde{k} = \frac{\sum_{i=1}^{N_\omega} |\hat{p}_{1,i}|}{\sum_{i=1}^{N_\omega} |\hat{v}_{1,i}|}.
\end{equation}
Equation~\eqref{eq:k_closed} admits a clear physical interpretation: the droop gain is chosen such that the aggregate magnitude of the converter's active-power response across all uncertainty sources matches the aggregate magnitude of its DC voltage excursion. This $\ell_1$-type aggregation treats each uncertainty source symmetrically, yields a single scalar gain compatible with the local droop implementation, and does not rely on assumptions about sign correlation between sources. When the system is driven by a dominant single source, \eqref{eq:k_closed} recovers the single-source expression \eqref{eq:k_single} as a special case. The ratio $\tilde{k}$ in~\eqref{eq:k_closed} is dimensionless: it is formed entirely from PCE coefficients of per-unit quantities and therefore reflects the intrinsic sensitivity structure of the system, independent of any specific base. To deploy the gain in the converter's local droop loop, this dimensionless ratio is mapped onto the engineering droop scale by a single zone-specific normalization constant $\alpha_z$:
\begin{equation}
\label{eq:k_norm}
k_{\text{opt}} = \alpha_z \cdot \tilde{k},
\end{equation}
where $\alpha_z$ is calibrated once per operating zone such that the median $k_{\text{opt}}$ across a small number of representative cases matches a reference engineering value $k_{\text{base}}= 20\,p.u.$ \cite{Beerten2013Analysis}. 

With this normalization in place, the extracted $k_{\text{opt}}$ varies continuously with the stochastic operating point encoded in the PCE coefficients. As the wind forecast traverses different zones and the Beta parameters in~\eqref{eq:beta_ab} adjust accordingly, the sensitivity coefficients $\hat{p}_{1,i}$ and $\hat{v}_{1,i}$ evolve, yielding an adaptive droop gain that inherits the probabilistic voltage-security guarantee encoded in the SOPF chance constraints. The resulting coefficient is applied as the slope of the droop characteristic at the designated converter station and is refreshed at each dispatch interval.

\section{Case Studies}
\label{sec:results}

\subsection{Test System and Setup}
\label{subsec:test_system}
\begin{figure}
    \centering
    \includegraphics[width=0.85\linewidth]{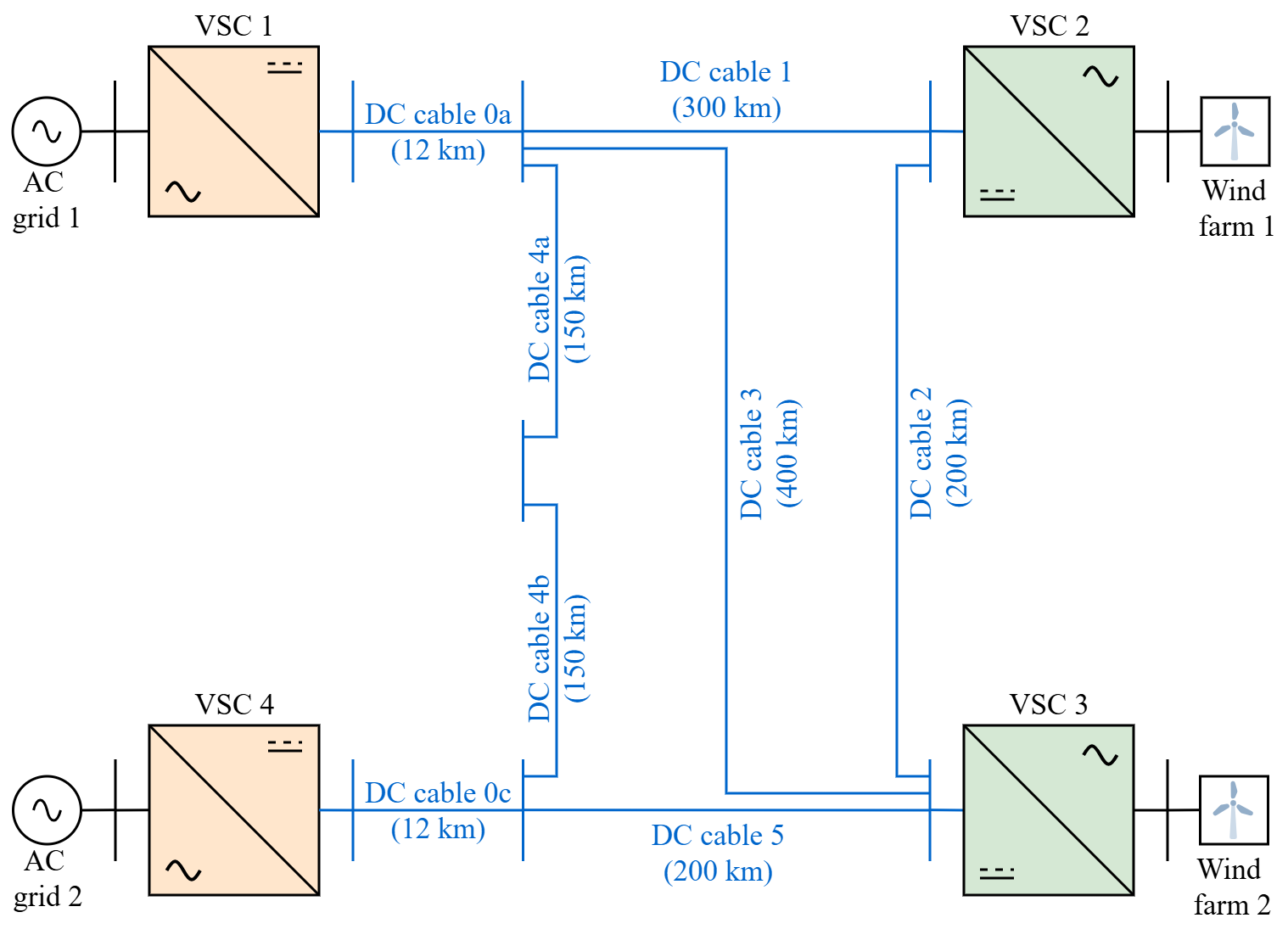}
    \caption{Configuration of a 4-terminal AC-MTDC system.}
    \label{fig:topology}
\end{figure}

The proposed framework is validated on a four-terminal HVDC grid coupled with two IEEE 9-bus AC subsystems and two offshore wind farms, as illustrated in Fig.~\ref{fig:topology}, which is the same test system used in our prior work~\cite{du2025opf}.  VSC 1 and VSC 4 interface the HVDC grid with the onshore AC grids, while VSC 2 and VSC 3 collect the offshore wind power from the two wind farms. VSC 1 is configured as the DC slack node and thus does not participate in droop regulation; VSC 2 and VSC 3, acting as offshore collector stations, operate in constant AC voltage and frequency mode to support the wind farms. The adaptive droop coefficient $k_{\text{opt}}$ derived from~\eqref{eq:k_norm} is therefore applied at VSC 4, which is the only onshore converter eligible for DC voltage droop control in this topology.

Each wind farm has a rated capacity of $p_{\max} = 1970.7 \, \mathrm {MW} $. A total of 90 forecast cases are generated, with 30 per operating zone (Low, Mid, High), as defined in Section~\ref{subsec:wind}. Within each case, the two wind farms draw their forecast levels from the same zone, but their forecast errors are treated as statistically independent, so that each farm contributes an independent stochastic germ $\xi_i$ to the PCE expansion. The realized wind power is then sampled from the fitted zone-specific Beta distribution, and the disturbance magnitude is defined as $\Delta = \|\mathbf{p}_{\text{real}} - \mathbf{p}_{\text{pred}}\|_2$, where $\mathbf{p}_{\text{real}}$ and $\mathbf{p}_{\text{pred}}$ stack the realized and forecast outputs of the two wind farms. 

The SOPF is formulated in the IV space with a confidence level $1-\epsilon = 0.95$ on the DC voltage limits $V_{\text{dc}} \in [0.9, 1.1]$\,p.u., and a PCE expansion degree of $d = 2$ is adopted, yielding six coefficients per random variable for the two-dimensional uncertainty. The resulting problem is solved by a primal-dual interior-point method in the PCE coefficient space. 

Three benchmark strategies are considered alongside the proposed adaptive scheme: two fixed-coefficient droop settings ($k = 20$\,p.u.\ and $k = 15$\,p.u., representing typical engineering choices), and a no-droop baseline ($k = 0$) in which VSC~4 operates in constant-power mode.

\subsection{Adaptive Droop Characteristics}
\label{subsec:droop_char}
\begin{figure}[t]
\centering
\includegraphics[width=0.95\columnwidth]{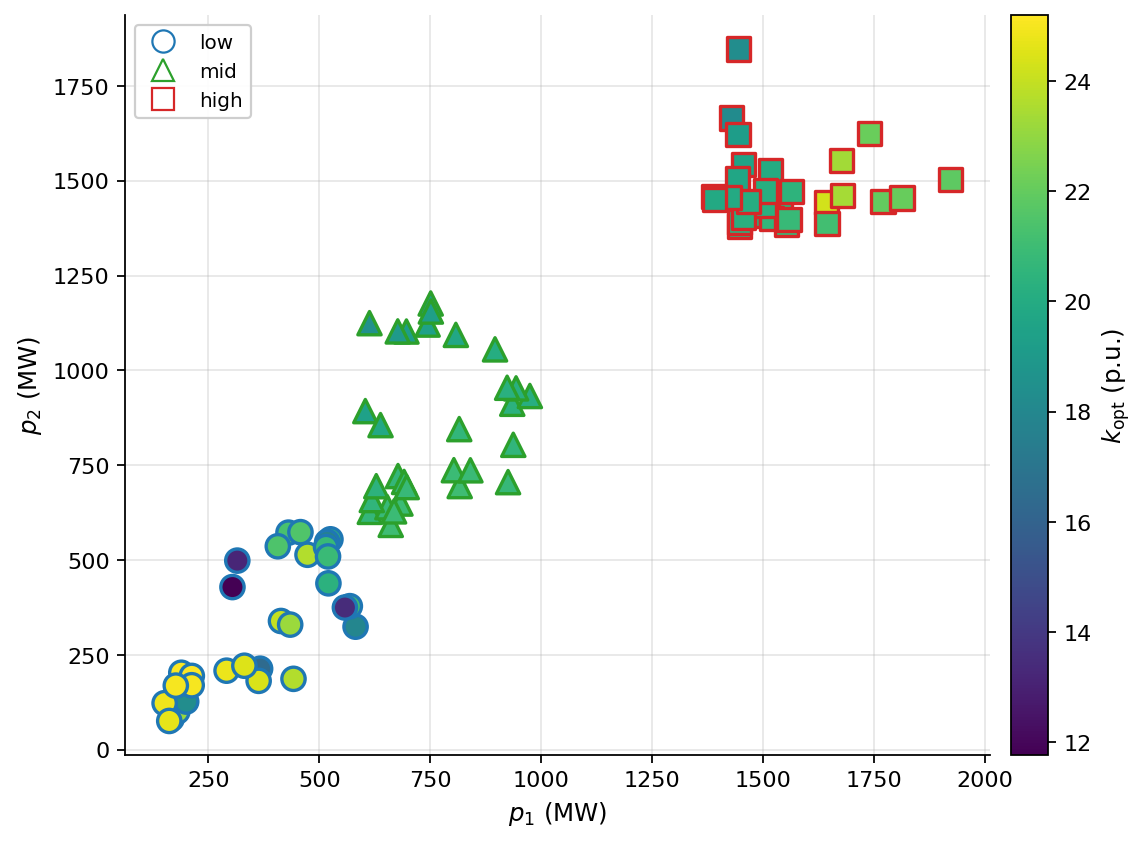}
\caption{Extracted adaptive droop coefficient $k_{\text{opt}}$ as a function of the two wind forecast inputs $(p_1, p_2)$ across all 90 operating cases. }
\label{fig:k_over_p1p2}
\end{figure}

Fig.~\ref{fig:k_over_p1p2} shows the extracted $k_{\text{opt}}$ as a function of the two forecast inputs $(p_1, p_2)$ across all 90 cases. The adaptive gain exhibits a clear dependence on the operating point: $k_{\text{opt}}$ tends to increase at low wind output and decrease toward the rated region, reflecting the converter's sensitivity to uncertainty, which changes significantly as the system approaches different portions of the power curve. This dependence cannot be captured by any fixed-coefficient droop scheme. 

\begin{figure}[t]
\centering
\includegraphics[width=0.87\columnwidth]{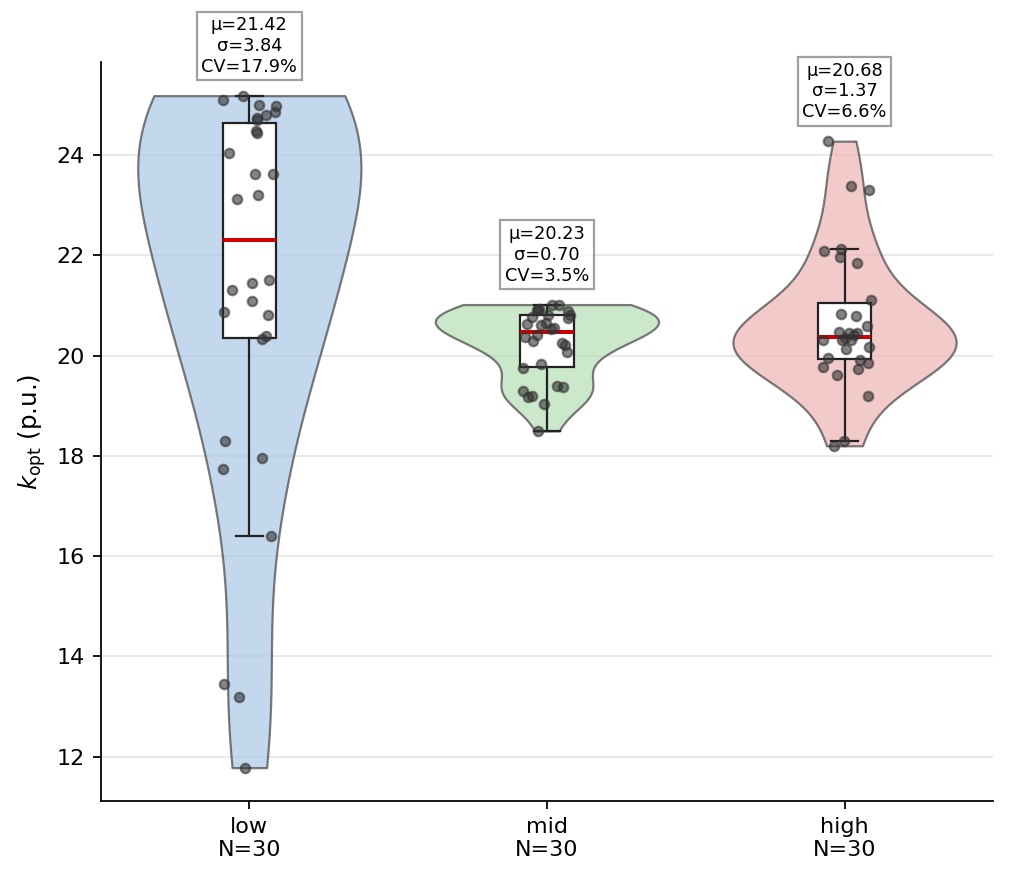}
\caption{Distribution of the extracted $k_{\text{opt}}$ within each operating zone (Low, Mid, High), 30 cases per zone. Annotations report the mean $\mu$, standard deviation $\sigma$, and coefficient of variation (CV) for each zone.}
\label{fig:k_violin}
\end{figure}

Fig.~\ref{fig:k_violin} further summarizes the $k_{\text{opt}}$ distribution within each operating zone. The Mid and High zones yield tightly concentrated distributions (CV of 3.5\% and 6.6\%, respectively), indicating that the proposed framework produces consistent droop settings under moderate and high wind conditions. The Low zone exhibits a broader spread (CV 17.9\%), which is consistent with the physical characteristics of the cut-in region. Small wind speed variations translate into large relative changes in converter loading and DC bus sensitivity, leading to a correspondingly wider range of optimal droop gains. Across zones, the medians of $k_{\text{opt}}$ range from 18 to 22\,p.u., confirming that the framework produces physically reasonable droop settings while continuously adapting to the stochastic operating point.

\subsection{Performance Under Disturbance}
\label{subsec:performance}

To quantify the operational benefit of the proposed adaptive droop in the presence of forecast errors, we evaluate the active-power tracking error of the droop-controlled converter VSC 4. For each case, the SOPF is solved on the basis of the wind forecast $\mathbf{p}_{\text{pred}}$, and the converter set-points $(P_{\text{VSC4}}^{\text{ref}},\, V_{\text{VSC4}}^{\text{ref}})$ together with the corresponding adaptive droop coefficient $k_{\text{opt}}$ dispatched to VSC 4. When the realized wind injections $\mathbf{p}_{\text{real}}$ subsequently materialize, the actual DC bus voltage settles at $V_{\text{VSC4}}^{\text{real}}$, and the local droop law produces the actual converter power
\begin{equation}
\label{eq:p_droop}
P_{\text{VSC4}}^{\text{droop}} = P_{\text{VSC4}}^{\text{ref}} - k_{\text{opt}}\,(V_{\text{VSC4}}^{\text{real}} - V_{\text{VSC4}}^{\text{ref}}).
\end{equation}

The corresponding ideal target, denoted $P_{\text{VSC4}}^{\text{truth}}$, is obtained by re-solving a deterministic AC--DC power flow with $\mathbf{p}_{\text{real}}$ as the injection and represents the converter operating point that the system would assume under perfect knowledge of the realization. The droop tracking error is then defined as
\begin{equation}
\label{eq:dev}
\Delta P = \left| P_{\text{VSC4}}^{\text{droop}} - P_{\text{VSC4}}^{\text{truth}} \right|,
\end{equation}
which directly measures how closely the local droop response approximates the ideal post-realization dispatch under wind forecast uncertainty. A smaller $\Delta P$ indicates that the converter, driven only by local DC voltage feedback through its droop characteristic, ends up at an operating point closer to the global optimum that would be reached by a re-optimization with full knowledge of the realized wind. The same metric is computed for the three benchmark strategies by replacing $k_{\text{opt}}$ with the corresponding fixed coefficient ($k=20$, $k=15$) or with $k = 0$ for the no-droop case.

\begin{figure*}[t]
\centering
\begin{subfigure}[t]{0.32\textwidth}
  \centering
  \includegraphics[width=\textwidth]{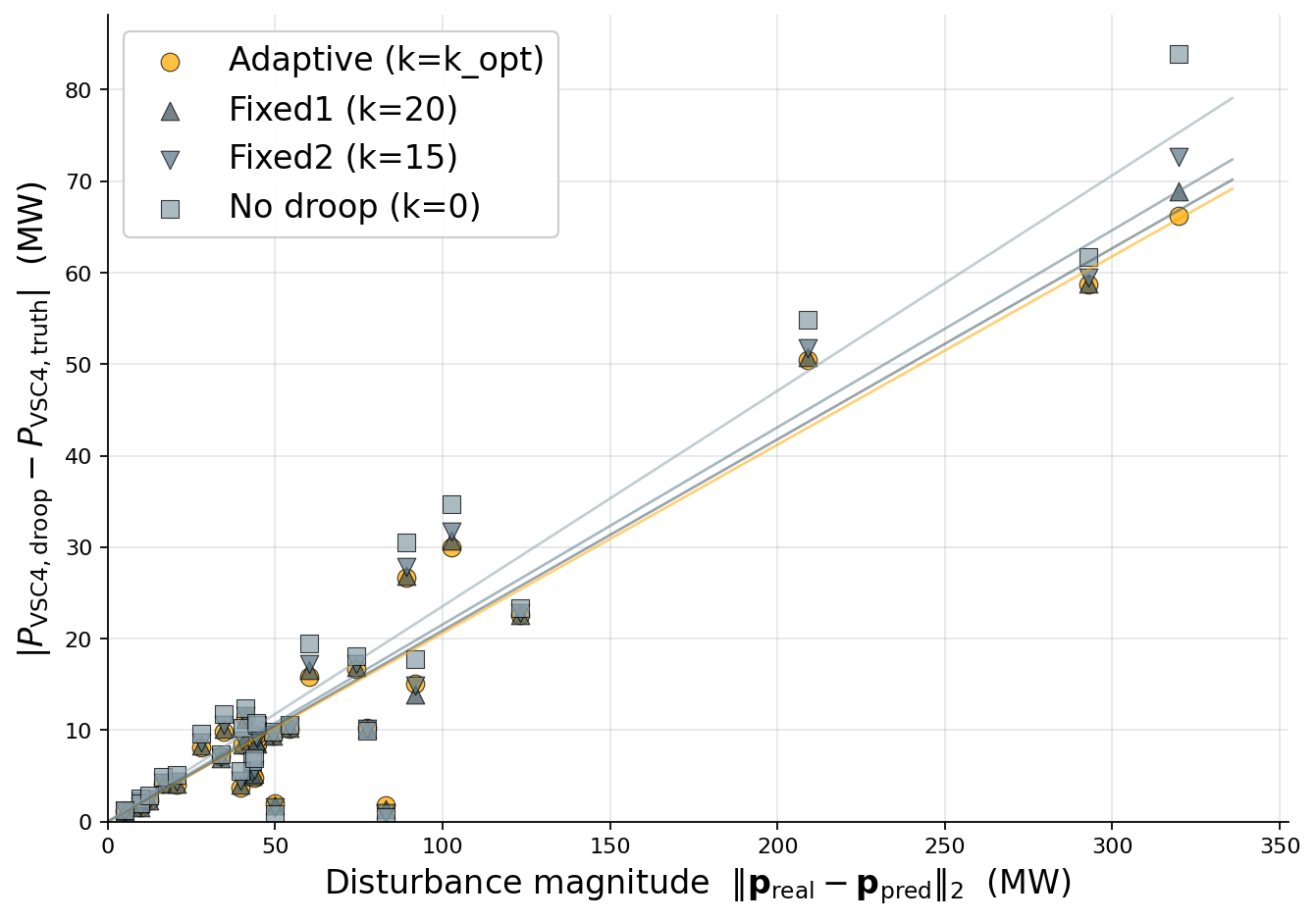}
  \caption{Low band}
  \label{fig:perf_low}
\end{subfigure}
\hfill
\begin{subfigure}[t]{0.32\textwidth}
  \centering
  \includegraphics[width=\textwidth]{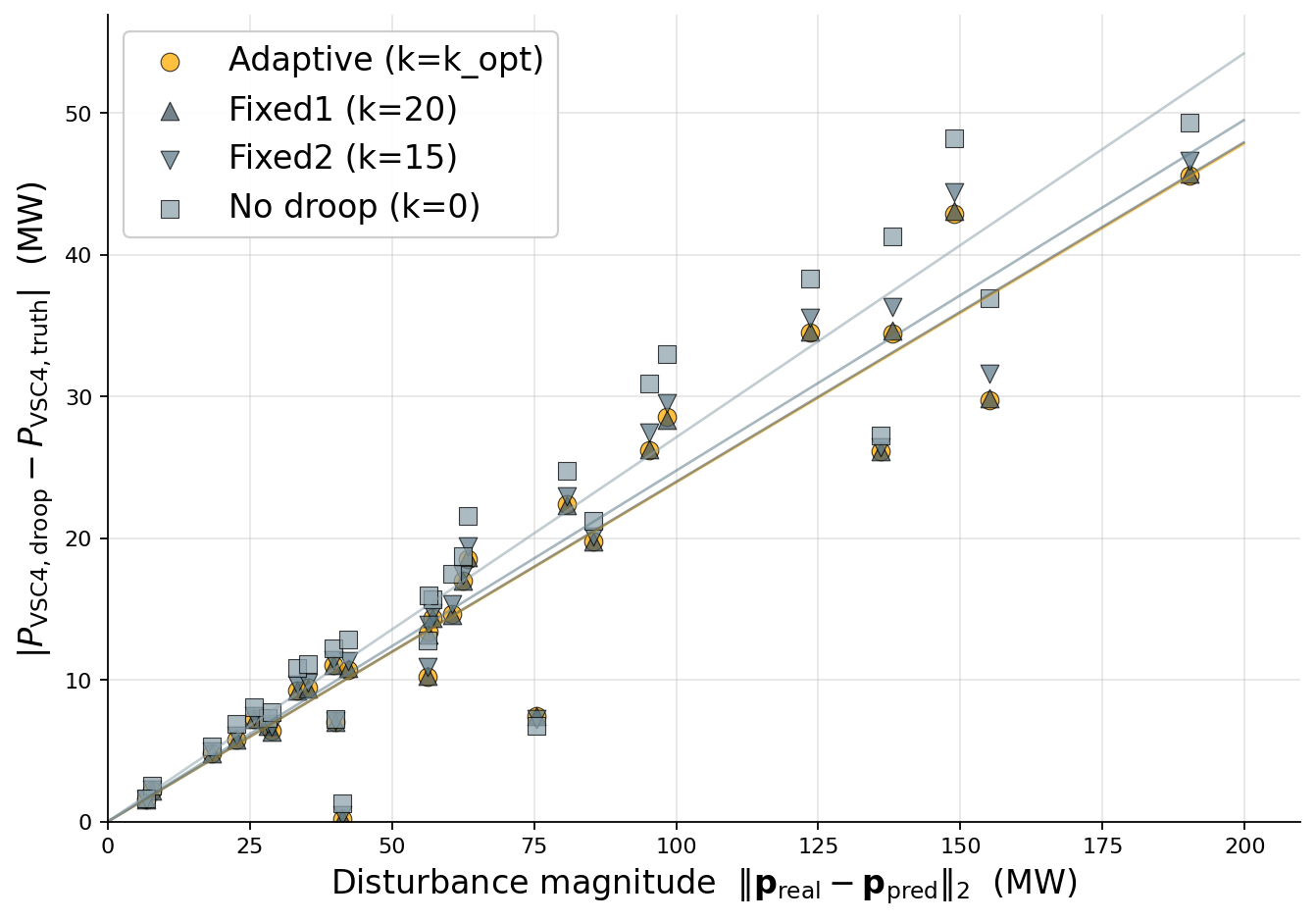}
  \caption{Mid band}
  \label{fig:perf_mid}
\end{subfigure}
\hfill
\begin{subfigure}[t]{0.32\textwidth}
  \centering
  \includegraphics[width=\textwidth]{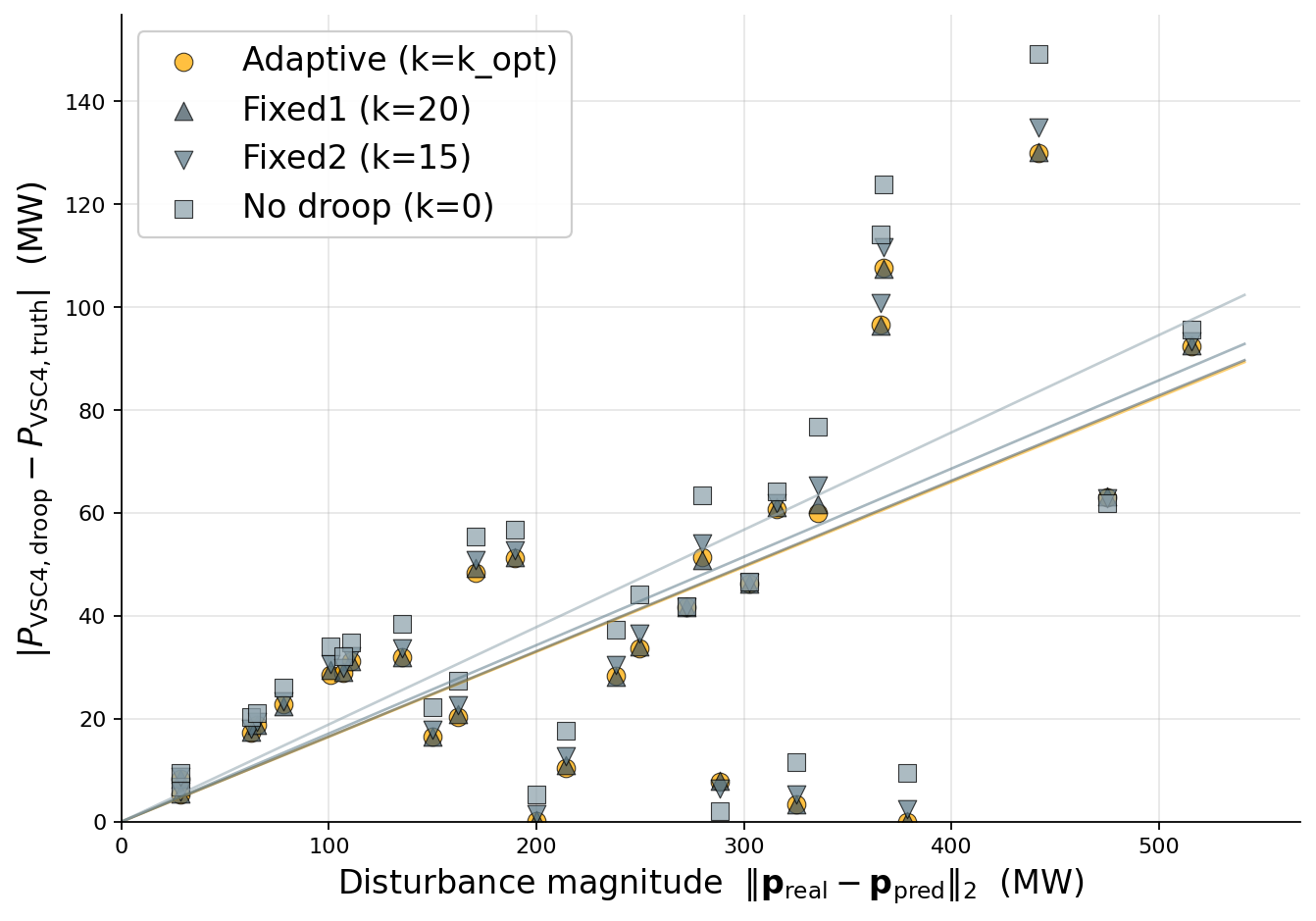}
  \caption{High band}
  \label{fig:perf_high}
\end{subfigure}
\caption{VSC 4 active-power tracking error $\Delta P$ as a function of the disturbance magnitude $\Delta = \|\mathbf{p}_{\text{real}} - \mathbf{p}_{\text{pred}}\|_2$, under four control strategies: adaptive $k_{\text{opt}}$, fixed $k=20$ and $k=15$, and no droop.}
\label{fig:perf_disturbance}
\end{figure*}

Fig.~\ref{fig:perf_disturbance} reports $\Delta P$ as a function of the disturbance magnitude $\Delta$ across the Low, Mid, and High bands, together with linear trend lines fitted to each control strategy. The trend lines provide a compact summary of how each strategy's tracking error grows with the disturbance. Across all three bands the adaptive scheme exhibits the smallest slope, followed by $k=20$, then $k=15$, with the no-droop baseline exhibiting the steepest growth. The ordering is consistent with the underlying design logic: under small disturbances all droop-based schemes perform comparably, as the linearized converter response is insensitive to the specific droop gain, but as $\Delta$ grows the fixed settings either over- or under-react to the DC voltage excursions, whereas $k_{\text{opt}}$ tracks the stochastic operating point through the PCE coefficients and maintains the tightest response. The separation between the adaptive and no-droop trend lines widens visibly from the Low band to the High band, confirming that the benefit of adaptive droop regulation is most pronounced under severe wind deviations.

Table~\ref{tab:performance} quantifies the per-case winner statistics together with the mean active-power tracking error across all strategies. The adaptive scheme attains the lowest deviation in 67\%--70\% of cases across all three disturbance bands, compared with only 23\%--30\% for the best fixed-coefficient setting ($k=20$); the fixed $k=15$ setting never attains the minimum, and the no-droop baseline prevails only in isolated outliers. The mean-error advantage over no-droop operation grows with disturbance magnitude and becomes most pronounced in the High band, where the adaptive scheme reduces the mean tracking error by 13.9\% (from 45.01 to 38.76\, MW), consistent with the widening trend-line separation in Fig.~\ref{fig:perf_disturbance}. Against the fixed $k=20$ baseline, the mean-error improvement appears modest, reflecting the fact that this value happens to lie close to the typical $k_{\text{opt}}$ in the Mid zone (Fig.~\ref{fig:k_violin}) and is therefore a favorable fixed choice a posteriori. Such a value, however, cannot be known in advance without extensive system-specific tuning, whereas the proposed framework automatically selects it from the PCE solution and continuously adjusts it with the operating point. The consistently higher winner rate of the adaptive scheme confirms that this adjustment produces a reliable per-case advantage rather than a coincidental average benefit.

\begin{table}[t]
\centering
\caption{Performance comparison across disturbance bands.}
\label{tab:performance}
\renewcommand{\arraystretch}{1.15}
\setlength{\tabcolsep}{4pt}
\begin{tabular}{llcccc}
\toprule
Band & Metric & Adaptive & $k=20$ & $k=15$ & No droop \\
\midrule
\multirow{2}{*}{Low}
  & Winner rate      & \textbf{67\%} & 23\% & 0\% & 10\% \\
  & Mean err.\ (MW)  & \textbf{14.26} & 14.37 & 14.83 & 16.22 \\
\midrule
\multirow{2}{*}{Mid}
  & Winner rate      & \textbf{67\%} & 30\% & 0\% & 3\% \\
  & Mean err.\ (MW)  & \textbf{16.28} & 16.30 & 16.86 & 18.52 \\
\midrule
\multirow{2}{*}{High}
  & Winner rate      & \textbf{70\%} & 23\% & 0\% & 7\% \\
  & Mean err.\ (MW)  & \textbf{38.76} & 38.97 & 40.48 & 45.01 \\
\bottomrule
\end{tabular}
\end{table}

\section{Conclusion}
\label{sec:conclusion}

This paper fundamentally bridges the long-standing operational gap between probabilistic system-wide dispatch and local converter-level regulation. By abandoning rigid fixed-gain assumptions, we introduce an adaptive droop control mechanism driven by SOPF under non-Gaussian offshore wind uncertainty. The technical cornerstone of this approach lies in translating zone-partitioned and Beta-distributed wind forecast errors into an interpretable droop coefficient using PCE sensitivities, bypassing explicit Jacobian formation and inversion. Case studies on a hybrid AC-HVDC system confirm the practical value of this approach: the derived adaptive gain dynamically adjusts to the prevailing stochastic operating point rather than relying on a nominal condition. By actively suppressing active-power tracking deviations under volatile conditions, the proposed control achieves a substantial reduction in mean error over a no-droop baseline. This ultimately guarantees that local converter responses are no longer heuristically tuned, but are fundamentally rooted in economic efficiency and probabilistic security. Future work will focus on extending this framework to incorporate $N-1$ security contingencies and validating the control performance through hardware-in-the-loop experiments.

\FloatBarrier
\bibliographystyle{IEEEtran}
\bibliography{manual.bib} 

\end{document}